\documentclass[prc,amsmath,twocolumn,superscriptaddress,preprintnumbers]{revtex4}

\usepackage{gensymb}
\usepackage{graphicx,color}
\usepackage{amssymb}
\usepackage{enumerate}
\usepackage{verbatim}
\usepackage{natbib}
\usepackage{amsmath}

\begin{document}
  \newcommand {\nc} {\newcommand}
  \nc {\Sec} [1] {Sec.~\ref{#1}}
  \nc {\IR} [1] {\textcolor{red}{#1}} 
  \nc {\IB} [1] {\textcolor{blue}{#1}}
  \nc {\CGMF}{$\mathtt{CGMF}$}

\title{Correlations Between Fission Fragment and Neutron Anisotropies in Neutron-Induced Fission}

\author{A.~E.~Lovell}
\email{lovell@lanl.gov}
\affiliation{Los Alamos National Laboratory, Los Alamos, NM 87545, USA}
\author{P.~Talou}
\affiliation{Los Alamos National Laboratory, Los Alamos, NM 87545, USA}
\author{I.~Stetcu}
\affiliation{Los Alamos National Laboratory, Los Alamos, NM 87545, USA}
\author{K.~J.~Kelly}
\affiliation{Los Alamos National Laboratory, Los Alamos, NM 87545, USA}

\date{\today}

%%%%%%%%%%%%%%%%%%%%%%%%%%%%%%%%%%%%%%%%%%%%%%%%%%%%%%%%%%%%%%%%%%%%%%%%%%%%%%%%%%%%%%%%%%%%%%%%%%%%%%%%%%%%%%%%%%%%%%%%%%%%%%%%%%%

\begin{abstract}
Several sources of angular anisotropy for fission fragments and prompt neutrons have been studied in neutron-induced fission reactions.  These include kinematic recoils of the target from the incident neutron beam and the fragments from the emission of the prompt neutrons, preferential directions of the emission of the fission fragments with respect to the beam axis due to the population of particular transition states at the fission barrier, and forward-peaked angular distributions of pre-equilibrium neutrons which are emitted before the formation of a compound nucleus.  In addition, there are several potential sources of angular anisotropies that are more difficult to disentangle:  the angular distributions of prompt neutrons from fully accelerated fragments or from scission neutrons, and the emission of neutrons from fission fragments that are not fully accelerated.  In this work, we study the effects of the first group of anisotropy sources, particularly exploring the correlations between the fission fragment anisotropy and the resulting neutron anisotropy.  While kinematic effects were already accounted for in our Hauser-Feshbach Monte Carlo code, {\CGMF}, anisotropic angular distributions for the fission fragments and pre-equilibrium neutrons resulting from neutron-induced fission on $^{233,234,235,238}$U, $^{239,241}$Pu, and $^{237}$Np have been introduced for the first time.  The effects of these sources of anisotropy are examined over a range of incident neutron energies, from thermal to 20 MeV, and compared to experimental data from the Chi-Nu liquid scintillator array.  The anisotropy of the fission fragments is reflected in the anisotropy of the prompt neutrons, especially as the outgoing energy of the prompt neutrons increases, allowing for an extraction of the fission fragment anisotropy to be made from a measurement of the neutrons.
\end{abstract}

\preprint{LA-UR-20-22603}

\maketitle
\section{Introduction}
\label{sec:intro}

%\begin{itemize}
	%\item Several experimental groups have shown that fission fragments are emitted anisotropically in the center of mass of the fissioning system (FissionTPC, etc. - I have a survey of this)
	%\item This emission was explained by A. Bohr - occurs due to the states in the parent nucleus that are populated
	%\item Neutrons that are emitted before fission should have a different angular dependence depending on whether they are emitted from the compound nucleus or emitted without the compound nucleus forming
	%\item Pre-fission neutrons were observed in $^{239}$Pu with the Chi-Nu array and shown to be in good agreement with the forward peaked FKK angular distributions
	%\item These two effects have been implemented into CGMF
%\end{itemize}

Being able to accurately and reliably model the fission process and the resulting observables is important for a variety of applications, including security, non-proliferation, energy, and fundamental science.  While microscopic models for this process, beginning from fundamental nucleon-nucleon interactions, are available \cite{Schunck2014,Bulgac2016}, more often these models are constructed based on a combination of macroscopic and microscopic theories \cite{Randrup2011,Moller2015}.  Still, at the moment, these methods are limited; they are computationally expensive, and the range of observables that can be calculated generally ends at the post-scission fission fragment yields.  These results can then be used as input for codes that calculate the prompt and delayed neutron and $\gamma$-ray observables.  

These codes (e.g, {\CGMF} \cite{CGMF,Becker2013}, $\mathtt{FREYA}$ \cite{FREYA1,FREYA2}, $\mathtt{HF^3D}/\mathtt{BeOH}$ \cite{BeoH}, $\mathtt{FIFRELIN}$ \cite{FIFRELIN}, $\mathtt{GEF}$ \cite{GEF}) employ a variety of physics models describing how the excitation and kinetic energies are shared between the light and heavy fission fragments, and determine neutron and $\gamma$-ray observables on either an event-by-event basis or as averages.  Event-by-event calculations allow for the determination of all observables, including correlations, as consistently as possible, from scission to prompt particle emissions, conserving energy, spin, and parity at every step of the decay.  %\IR{Not very informative:  However, to accurately describe the fission process and to compare to experiment, these models have to include as much of the physics as possible.}
One effect that has been studied over the years, but is not always fully included in these types of codes, is the anisotropic emission of fission fragments and neutrons during the decay.  There are several sources of this anisotropy:  fission fragment emission with respect to the incident beam, neutron emission from the fission fragments (including whether neutrons are emitted from the fully accelerated fragments), along with the emission of scission, multi-chance, and pre-equilibrium neutrons before fission.    

Although, there is no reason to expect that fission fragments should be emitted anisotropically with respect to the incoming neutron beam from the compound nucleus, many experimental studies from the 1950's through today have shown that there can be a large degree of anisotropy - defined as the ratio of the cross section at $0^\circ$ to the cross section at $90^\circ$.  The enhancement with respect to the beam axis can be as much as 30\% to 60\%.  There have, in particular, been several measurements of the anisotropy of actinides, including $^{235}$U \cite{Vorobyev2015,Ahmad1979,Brolley1955,Simmons1960,GeppertKleinrath2017}, $^{238}$U \cite{Vorobyev2015,Henkel1956,Brolley1954,Birgersson2009,Ryzhov2005,Vives2000,Simmons1960}, and $^{239}$Pu \cite{Leachman1965,Blumberg1959,Simmons1960}.  

This anisotropy was postulated, by A. Bohr \cite{Bohr1956}, to stem from the populated transition states at the fission barrier of the parent nucleus, which would account for the changing anisotropy as a function of incident neutron energy.  At low incident energies, few states are populated, resulting in an anisotropic distribution of fission fragments depending on the quantum numbers of the states that are populated.  As the incident energy increases, more states are available in the parent nucleus (and the angular distribution becomes more isotropic) until the second-chance fission channel opens up, at which point the $A$ nucleus is fissioning instead of the $A+1$ system, lowering the number of available states, increasing the anisotropy again.  Of course, the picture is more complicated than this since the opening of the second-chance fission channel leads to a possibility of fission from either the $A+1$ or $A$ compound nucleus.  A recent measurement of the anisotropy of $^{239}$Pu confirmed that the changes in anisotropy track the shape of the fission cross section \cite{GeppertKleinrath2017}, in agreement with many previous measurements that observed the same trends, e.g. \cite{Leachman1965,Blumberg1959,Simmons1960}. 

There have also been several studies to assess whether or not the prompt neutrons are emitted isotropically from the fission fragments, along with whether the neutrons are emitted from the fully accelerated fragments, both of which are commonly assumed in the fission fragment deexcitation codes mentioned above.  Often, differences between the measured and calculated neutron angular distributions are attributed to scission neutrons, those neutrons that are emitted during the rupture of the neck of the compound nucleus.  However, this interpretation does not take into account the assumptions in the model that could change the resulting calculations (such as that all prompt neutrons are emitted from the fully accelerated fragments).  The percentage of neutrons that are not emitted from fully accelerated fission fragments is somewhat of a controversy \cite{Bowman1962,Bowman1963,Kapoor1963,Skarsvag1963,Carjan2019}.

As the incident neutron energy increases, reaction channels beyond first-chance fission open.  Besides the scission neutrons that may be emitted, other pre-fission neutrons can be emitted before the compound nucleus fissions.  Most often, these pre-fission neutrons are emitted from the fully equilibrated compound nucleus.  However, for incident neutrons with high enough energy, $\gtrsim 12$ MeV, a neutron can be emitted with or without the compound nucleus forming.  When the compound nucleus does not form, the emitted neutrons are considered to be pre-equilibrium.  These pre-equilibrium neutrons have a different energy spectrum from the pre-fission neutrons coming from the compound system \cite{PreEqReactions} and have a forward-peaked angular distribution, more akin to the angular distribution of an inelastically scattered neutron than the isotropic distributions of the pre-fission neutrons emitted from the compound system \cite{Feshbach1980}.  Pre-equilibrium neutrons have been definitively observed and their angular distributions measured for the first time for neutron-induced fission of $^{239}$Pu \cite{Kelly2019}.

Recently, the fission fragment anisotropy and pre-equilibrium angular distributions have been implemented into the Monte Carlo Hauser-Feshbach code {\CGMF} \cite{CGMF,Becker2013}.  %\IR{In addition to describing the physics models as accurately as possible, these effects are crucial to being able to compare with experimental results, such as those from the Chi-Nu array at Los Alamos which measures the energy and angle of fission neutrons \cite{Haight2015,Devlin2018}.}
In this work, we describe the implementation and study of the correlations between the fission fragment and neutron directions.  This paper is organized in the following manner:  in Section \ref{sec:theory}, we discuss the underlying physics of the angular distributions as well as the {\CGMF} code and the implementation of these sources of anisotropy; in Section \ref{sec:results}, we present our results and compared to experimental data; and finally, we conclude in Section \ref{sec:conclusions}.

%%%%%%%%%%%%%%%%%%%%%%%%%%%%%%%%%%%%%%%%%%%%%%%%%%%%%%%%%%%%%%%%%%%%%%%%%%%%%%%%%%%%%%%%%%%%%%%%%%%%%%%%%%%

\section{Theory}
\label{sec:theory}

%\begin{itemize}
	%\item Fission fragment anisotropy (Bohr)
	%\item Pre-equilibrium angular distributions
	%	\item CGMF
	%	\item CGMF doesn't take into account:  non-isotropic distribution of neutrons from fission fragments, neutron emission from accelerating fragments, scission-neutrons, spin-dependence of neutron emission
%\end{itemize}

There are several physical processes that could lead to an anisotropic distribution of the prompt neutrons emitted during the fission process.  The following sections describe these processes along with the Monte Carlo Hauser-Feshbach code {\CGMF}, used here to calculate the decay of the fission fragments by prompt neutron and $\gamma$-ray emissions.

\subsection{Fission Fragment Anisotropy} 
\label{sec:FFA}

%In a na\"ive picture, there is no reason to assume that fission fragments from neutron-induced fission should have a preferential direction to decay with respect to the beam axis.  However, there have been many studies across multiple isotopes that show there is a preferential direction, and that this direction changes with the incident neutron energy.  It was in 1956 that A. Bohr \cite{Bohr1956} postulated that this preferential direction was due to the specific transition states that are populated at the outer saddle point.  

The first process we discuss is the fission fragment angular anisotropy coming from the population of the transition states \cite{Bohr1956}.  The angular distribution of the fission fragments can be described with the help of the Wigner $d$-matrices \cite{Wagemans}, defined as 
\begin{multline}
%\begin{split}
d^{J}_{M,K} (\theta) = \\
\sum \limits _n (-)^n \frac{[(J+M)!(J-M)!(J+K)!(J-K)!]^{1/2}}{(J-M-n)!(J+K-n)!(M-K+n)!n!} \\
\times \left ( \mathrm{cos}\frac{\theta}{2}\right )^{2J+K-M-2n} \left ( \mathrm{sin}\frac{\theta}{2}\right )^{2n+M-K},
%\end{split}
\end{multline}

\noindent where $J$ is the total angular momentum of the compound nucleus, $K$ is its projection on the fission axis, $M$ is its projection on the beam axis, and $\theta$ is the angle of the fission fragments with respect to the beam axis.  The angular distribution of the fission fragments in the lab frame would be 
\begin{equation}
\frac{d\sigma_f}{d\Omega} (J,K,M) = \frac{2J+1}{2} | d^J_{M,K} (\theta) |^2,
\end{equation}

\noindent if the fission only occurred through one of these $(J,K,M)$ states.  Instead, each state must be weighted by its population, $P_s(J,K,M)$, and the full angular distribution is given by
\begin{equation}
W(\theta) \equiv \frac{d\sigma_f}{d\Omega} = \sum \limits _{s(J,K,M)} P_s(J,K,M) \frac{2J+1}{2} | d^J_{M,K} (\theta) |^2.
\end{equation}

\noindent The weight of each state is often extracted from experimental measurements, e.g. \cite{GeppertKleinrath2017}, but these weights are connected to the fission cross section, $\sum P_s = \int \frac{d\sigma_f}{d\Omega}d\Omega$.  The anisotropy of the fission fragments is defined as the ratio of the angular distribution at 0$^\circ$ to the angular distribution at 90$^\circ$ relative to the beam axis, often written as $W(0^\circ)/W(90^\circ)$.

\subsection{Pre-equilibrium Neutron Angular Distribution}
\label{sec:PEA}

When the incident neutron energy is high enough, pre-equilibrium neutrons can be emitted before the compound system equilibrates.  Both the energy spectrum and angular distributions of these neutrons are significantly different from those of the neutrons emitted from the fully equilibrated compound nucleus.  Pre-equilibrium angular distributions can be well described by the quantum mechanical theory of Feshbach, Kerman, and Koonin (FKK) \cite{Feshbach1980}.  Here, the angular distributions of the emitted particles can be separated into multi-step direct and multi-step compound parts.  The pre-equilibrium neutrons are described by the multi-step direct part of the calculation, similar to inelastic scattering.  

Recently, pre-equilibrium angular distributions were measured from $^{239}$Pu(n,f) for the first time \cite{Kelly2019}.  These neutrons were shown to be first emitted around an incident neutron energy of 12 MeV and were significantly forward peaked.  These forward-peaked angular distributions were well described by the FKK theory, along with Kalbach-Mahn systematics \cite{Kalbach1981}, which represents the shape of the angular distributions as a sum of Legendre polynomials. 

\subsection{Other Sources of  Anisotropy}
\label{sec:otherA}

Another source of angular anisotropy for the fission fragments and the emitted neutrons is from the recoil of the compound system due to momentum conservation between the neutron beam and the fissile target.  Although the neutrons are very light compared to the heavy target, this effect is non-negligible.  The recoil had already been included in {\CGMF}, and the resulting effect on the fission fragment and neutron anisotropy will be presented in Sec. \ref{sec:results}.  Similarly, there is also a small amount of recoil to the fission fragments from the emission of the prompt neutrons; this kinematic effect was also already included in {\CGMF}.

There are several other potential sources of anisotropy that depend on how the neutrons are emitted from the fission fragments.  In this work, we assume that the prompt neutrons are emitted isotropically from the fission fragments in their center-of-mass frame.  In addition, we assume that the neutrons are emitted from the fully accelerated fission fragments.  Anisotropic angular distributions of the neutrons in the center-of-mass frame of the fragments could either raise or lower the resulting neutron anisotropy depending on the distribution, while neutrons emitted before the fragments are fully accelerated would reduce the anisotropy of the neutrons with respect to the fragments as the neutrons would receive a smaller kinematic boost from the fragments.  Both of these effects should be investigated and then taken into account, however, that is beyond the scope of this current work.

\subsection{CGMF}
\label{sec:CGMF}

The fission fragment anisotropy and pre-equilibrium angular distributions have recently been implemented into the Monte Carlo Hauser-Feshbach code, {\CGMF} \cite{CGMF,Becker2013}, developed at Los Alamos National Laboratory (LANL).  As input, {\CGMF} requires yields in mass, charge, total fission fragment kinetic energy, spin, and parity, $Y(A,Z,TKE,J,\pi)$, for the fission reaction of interest, either spontaneous fission or neutron-induced fission for incident energies from thermal to 20 MeV.  The fission fragments are sampled from this initial distribution.  The total excitation energy, TXE, defined through the $Q$-value as TXE = $Q$ - TKE, is calculated and shared between the two complementary fragments based on a ratio of temperatures, currently fit to reproduce the average neutron multiplicity as a function of mass and does not include a dependence on the incident neutron energy.  Each fragment decays through the emission of prompt neutrons and $\gamma$ rays according to the Hauser-Feshbach statistical theory \cite{Feshbach1980}.  At each step in the decay, the probability for the emission of a prompt neutron is determined from the transmission coefficient calculated from an optical potential, and the $\gamma$ emission probability is calculated from the $\gamma$-ray strength function.  Energy, spin, and parity are conserved during the full decay path.

{\CGMF} records the initial conditions of the fragments along with the energies and momenta of the fragments, neutrons, and $\gamma$ rays for each simulated fission event.  This allows us to reconstruct average quantities, such as average neutron energy, photon energy, and number of particles, as well as distributions, such as energy spectra and multiplicity distributions.  In addition, correlations between all observables can be constructed.  Numerous examples of the prompt quantities that have been studied with {\CGMF} can be found in \cite{Talou2018,Stetcu2014,Marcath2018}.

As mentioned previously, {\CGMF} now takes into account the angular distributions of fission fragments and of the pre-equilibrium neutrons.  As far as we are aware, this is the first time that these two effects have been included into a fission decay code such as this.  In the sections below, we describe how these angular distributions were parametrized and included within {\CGMF}.

%%%%%%%%%%%%%%%%%%%%%%%%%%%%%%%%%%%%%%%%%%%%%%%%%%%%%%%%%%%%%%%%%%%%%%%%%%%%%%%%%%%%%%%%%%%%%%%%%%%%%%%%%%%%%%

%\section{Experimental Details}
%\label{sec:experiment}

% In here, we would put details of Chi-Nu, the experiment, and the analysis, especially if things are non-standard due to looking at counts and not just shape measurements.

%%%%%%%%%%%%%%%%%%%%%%%%%%%%%%%%%%%%%%%%%%%%%%%%%%%%%%%%%%%%%%%%%%%%%%%%%%%%%%%%%%%%%%%%%%%%%%%%%%%%%%%%%%%%%%

%\section{Numerical Details}
%\label{sec:numbers}

%\begin{itemize}
%	\item Process of fitting the experimental anisotropy for three target nuclei using cubic splines
%	\item Process of fitting the FKK calculations for the pre-fission neutron distributions - polynomials as a function of incident energy and excitation energy (same fit for all nuclei)
%\end{itemize}

%In order to correctly sample the anisotropic angular distribution for the fission fragment and pre-equilibrium neutrons, these distributions had to be included in {\CGMF}.  %The Monte Carlo nature of {\CGMF} makes it already computationally expensive, so instead of directly implementing models for these anisotropies, experimental and theoretical distributions were parameterized and sampled within the code.  
%These parameterizations are described in the following section.

\subsubsection{Fission Fragment Anisotropy}

\begin{figure}
	\centering
	\includegraphics[width=0.45\textwidth]{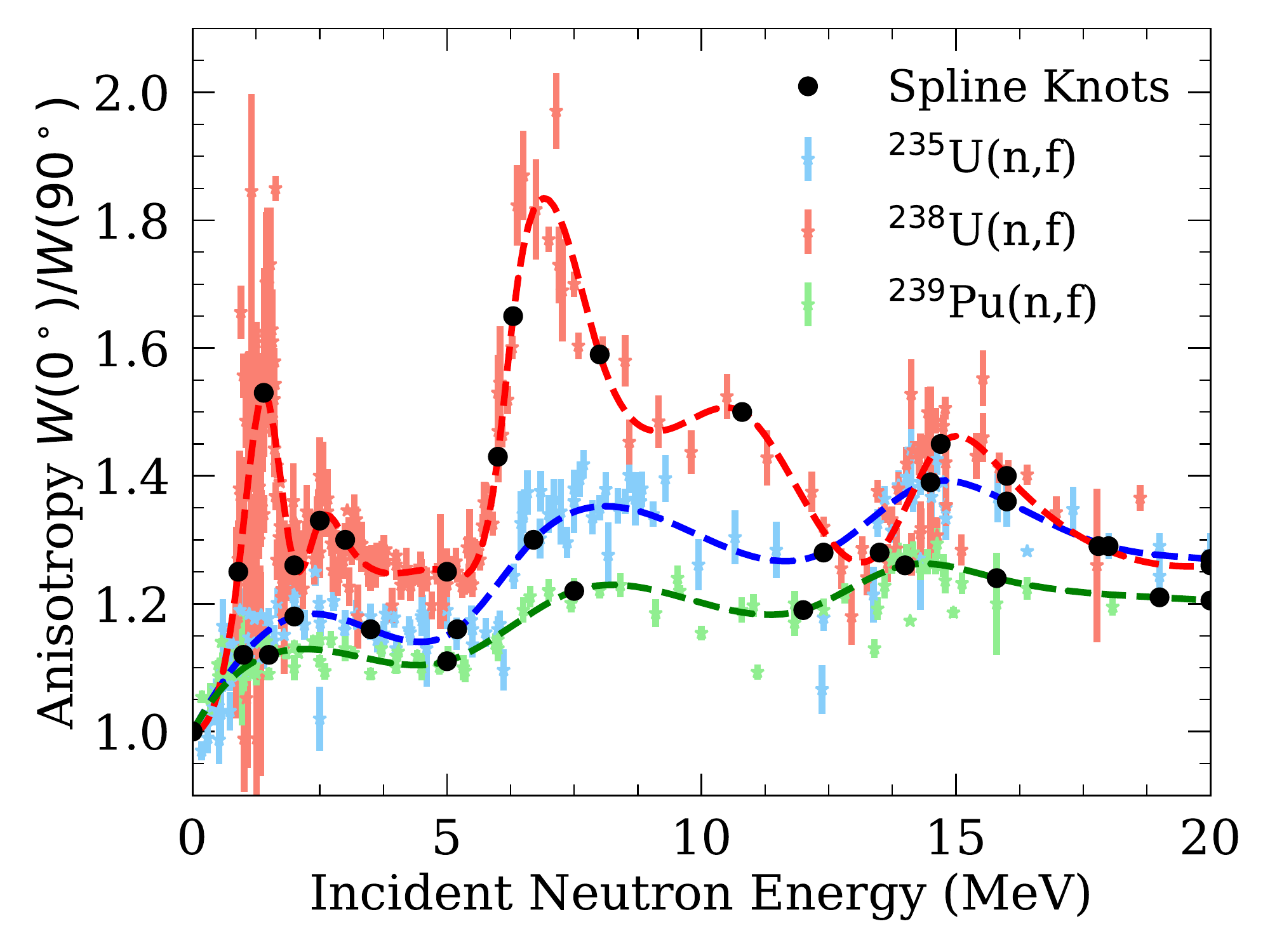}
	\caption{(color online) Fission fragment anisotropy as a function of incident neutron energy.  Stars with error bars show the data, black circles are the knots that anchor the cubic splines, given by the dashed lines of the corresponding color, $^{235}$U(n,f) in blue, $^{238}$U(n,f) in red, and $^{239}$Pu(n,f) in green.  These represent some of the fits now included in {\CGMF}.}
	\label{fig:Splines}
\end{figure}

The fission fragment anisotropy, $W(0^\circ)/W(90^\circ)$, was introduced in {\CGMF} for all neutron-induced reactions available.  For most of these nuclei, there are several experimental data sets available, shown in Fig. \ref{fig:Splines} for $^{235}$U, $^{238}$U, and $^{239}$Pu as the colored stars.  For each nucleus, these data were fit using cubic splines, shown as the correspondingly colored dashed lines of Fig. \ref{fig:Splines}.  The knots of the splines, black circles, were adjusted to reproduce the features within the experimental data.  Cubic splines fits were also constructed for the anisotropy of $^{233,234}$U, $^{237}$Np and $^{241}$Pu.  However, for the $^{241}$Pu target, data were only available for incident neutron energies below 8 MeV, and because these data have the same shape and magnitude as those for $^{239}$Pu, the anisotropy as a function of incident neutron energy for $^{241}$Pu was taken to be the same as that of $^{239}$Pu, where the data extend up to 20 MeV.  For implementation into {\CGMF}, these fits are tabulated in steps of 0.1 MeV incident neutron energy.  This energy is then matched with the incident neutron energy for each {\CGMF} calculation.  The uncertainty introduced from this rounding is less than the accuracy of the resulting anisotropy determined from a fit to the Monte Carlo results.

The angle, $\theta$, with respect to the incident neutron beam axis of the light fission fragment is then sampled from the distribution
\begin{equation}
P(\mathrm{cos}(\theta)) \propto 1+\left (\frac{W(0^\circ)}{W(90^\circ)}-1 \right )\mathrm{cos}^2(\theta),
\label{eq:FFA}
\end{equation}

\noindent in the center-of-mass frame of the compound system.  When $W(0^\circ)/W(90^\circ)=1$, $\mathrm{cos}(\theta)$ is sampled isotropically.  In all cases, the angle around the beam axis, $\phi$, is sampled isotropically.  The direction of the heavy fragment is determined from the back-to-back momentum conservation of the fragments in the center-of-mass.

This method, however, does not rigorously take into account the opening of multi-chance fission channels.  Ideally, when a fissioning nucleus emits one or more pre-fission neutrons, the anisotropy of the $A_c-\nu$ system - where $\nu$ is the number of pre-fission neutrons emitted - should be considered in Eq. (\ref{eq:FFA}), instead of the anisotropy of the $A_c$ system.  From the measurement of the fission fragment anisotropy, it is impossible to disentangle the joint effect of the anisotropy of each fissioning system without further modeling.

\subsubsection{Pre-equilibrium Neutron Angular Distributions}

The pre-equilibrium neutron angular distributions were fit to theoretical calculations from the FKK model \cite{Feshbach1980,TKcomm}.  These calculations were performed on a two-dimensional grid of incident, $E_\mathrm{inc}$, and excitation, $E^*$, energies and were individually fitted assuming Kalbach systematics \cite{Kalbach1988}, 
\begin{equation}
\frac{d\theta}{d\Omega}(E_\mathrm{inc},E^*) = a(E_\mathrm{inc},E^*)\mathrm{sinh}(\mathrm{cos}\theta) + b(E_\mathrm{inc},E^*).
\end{equation}

\noindent Here, both $a$ and $b$ were taken to be polynomial in $E_\mathrm{inc}$ and $E^*$.  This parameterization allows an angular distribution to be calculated for any incident neutron energy up to 20 MeV.  The resulting distributions are then normalized to form a probability distribution function and are sampled when a pre-equilibrium neutron is emitted.  These distributions were calculated for $^{239}$Pu but are taken to be the same for all targets.  For the actinides considered in {\CGMF}, these angular distributions calculated with the FKK model are almost unchanged with the addition or removal of a few nucleons.

The emitted neutrons are then sampled from a cumulative distribution function, CDF.  The CDF was constructed for 50 values of $\mathrm{cos}(\theta) \in [-1,1]$, and points in between were interpolated linearly.  Here, 50 values of $\mathrm{cos}(\theta)$ are enough for converged results, and this grid and interpolation does not introduce any uncertainty compared to a finer grid.  The angular distributions of the fission fragments were sampled in the same way.

\subsubsection{Extracting the Anisotropy}
\label{sec:extractA}

Because {\CGMF} calculates the fission events in a Monte Carlo fashion, for each run of {\CGMF}, we end up with a list of properties for each of the fission fragments and associated neutrons and $\gamma$ rays.  To extract the neutron anisotropy, we bin the neutron angles and fit the resulting distribution to Legendre polynomials,
\begin{equation}
W(\mathrm{cos}\theta|\textbf{a}) = a_0P_0 (\mathrm{cos}\theta) + a_1P_1 (\mathrm{cos}\theta) + a_2P_2 (\mathrm{cos}\theta),
\end{equation}

\noindent where $P_0$, $P_1$, and $P_2$ are the first three Legendre polynomials, and $\textbf{a}=[a_0,a_1,a_2]$ are the associated weights of each polynomial.  From each fit, we get best-fit values, $a_0^*$, $a_1^*$, and $a_2^*$, along with a covariance matrix between the three fit parameters.  To quantify the uncertainty coming from the polynomial fit, we draw samples from a multivariate Gaussian distribution centered at the best fit with the associated covariance matrix.  The anisotropy of the neutrons is then defined as $W(1|\textbf{a}^*)/W(0|\textbf{a}^*)$, and 95\% confidence bands are constructed from the samples from the covariance matrix.  This procedure is used to mitigate the lack of statistics when looking at specific neutron energies.  The same procedure is followed to extract the anisotropy of the fission fragments from the {\CGMF} calculations.  

%%%%%%%%%%%%%%%%%%%%%%%%%%%%%%%%%%%%%%%%%%%%%%%%%%%%%%%%%%%%%%%%%%%%%%%%%%%%%%%%%%%%%%%%%%%%%%%%%
\section{Results and Discussion}
\label{sec:results}

%\begin{itemize}
	%\item Have results for three systems - $^{239}$Pu, $^{235}$U, $^{238}$U
	%\item Differences between isotropic sampling everywhere, only including the fission fragment anisotropy, including fission fragment and pre-equilibrium neutron anisotropy (for the last, all neutrons vs. just pre-fission neutrons)
	%\item Effect on neutrons as a function of neutron threshold energy
	%\item Neutron anisotropy as a function of incident neutron energy compared to the anisotropy of the fission fragments 
%\end{itemize}

\begin{figure}
\centering
\includegraphics[width=0.45\textwidth]{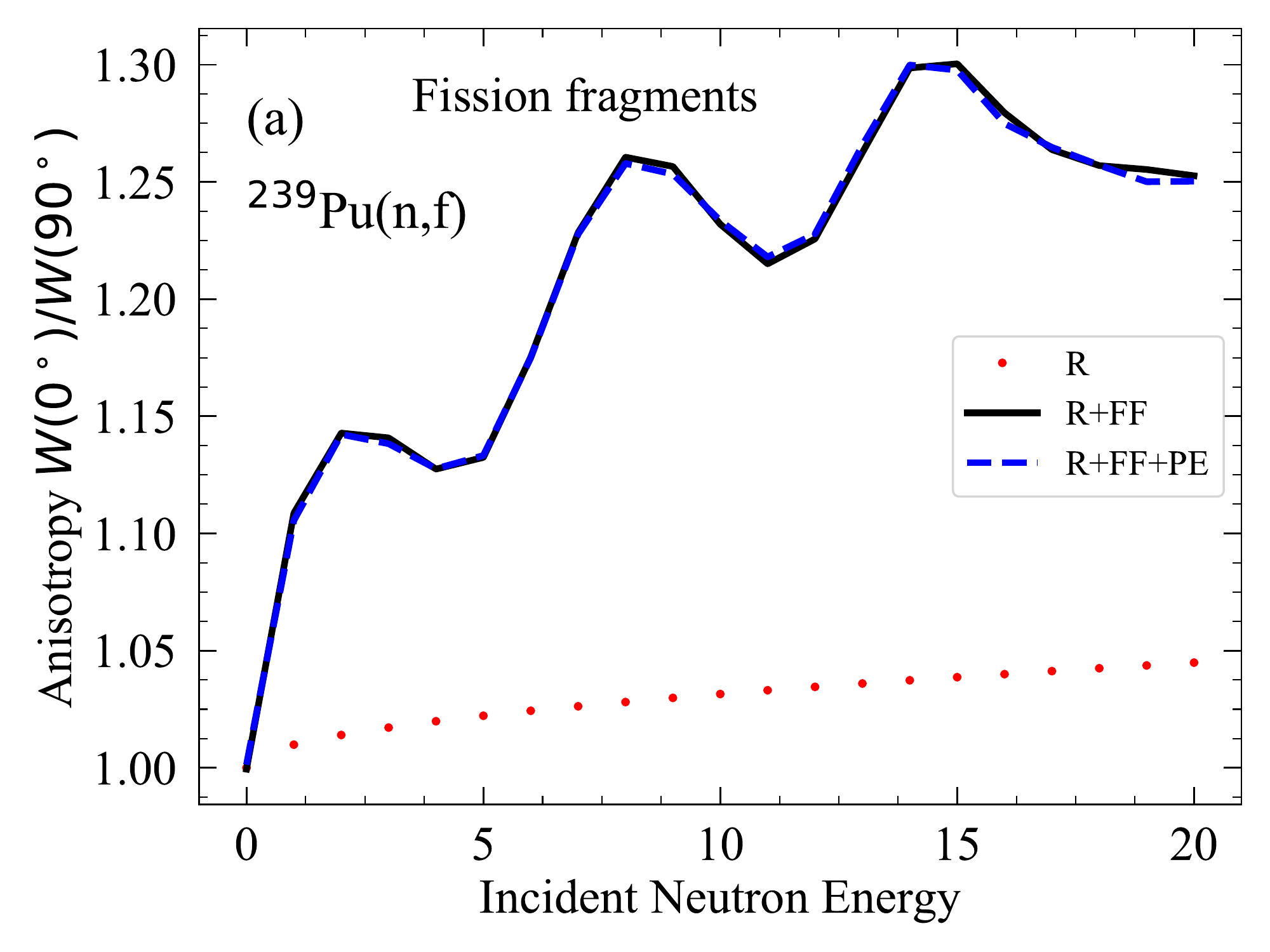} \\
\includegraphics[width=0.45\textwidth]{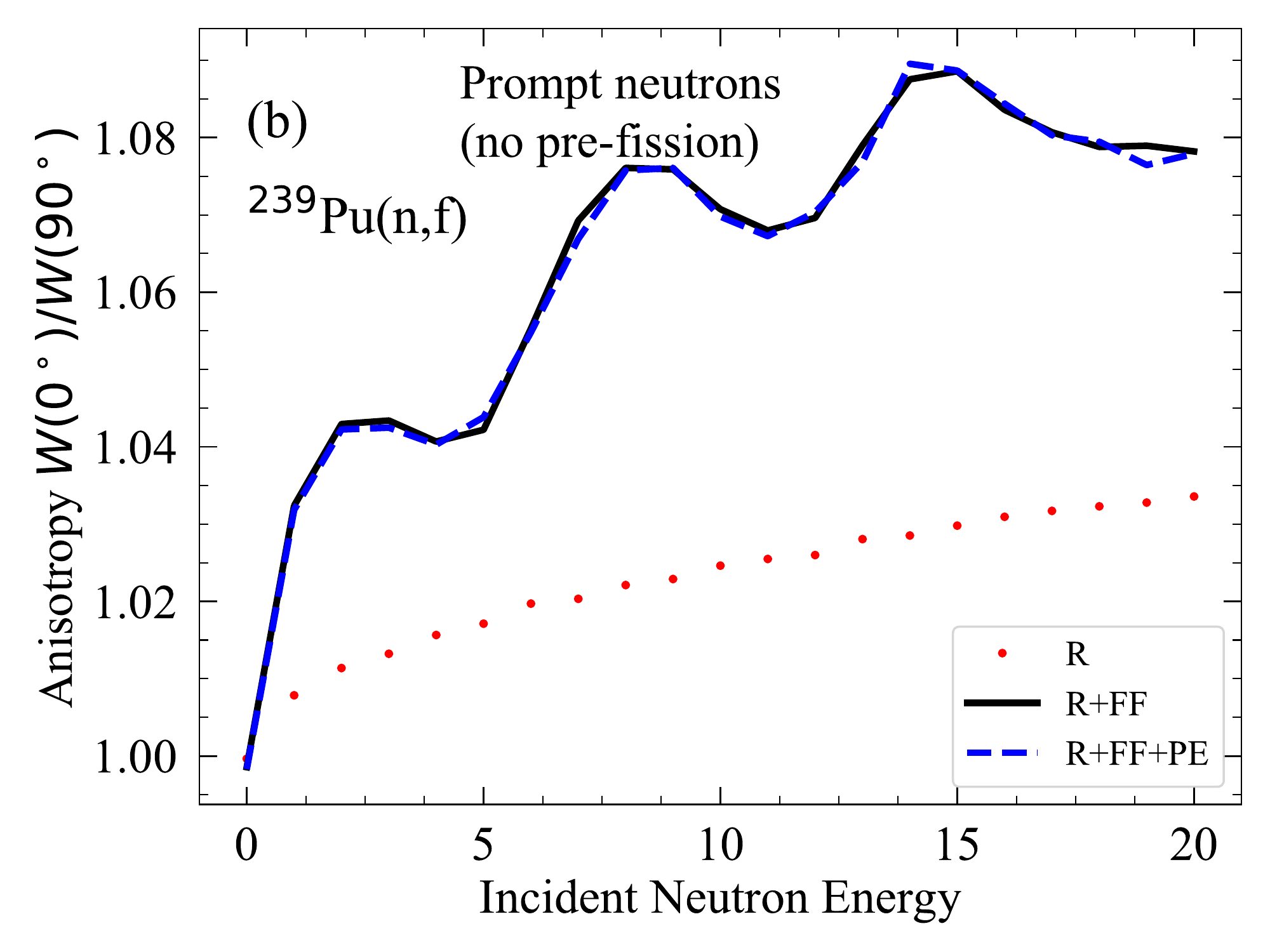} \\
\includegraphics[width=0.45\textwidth]{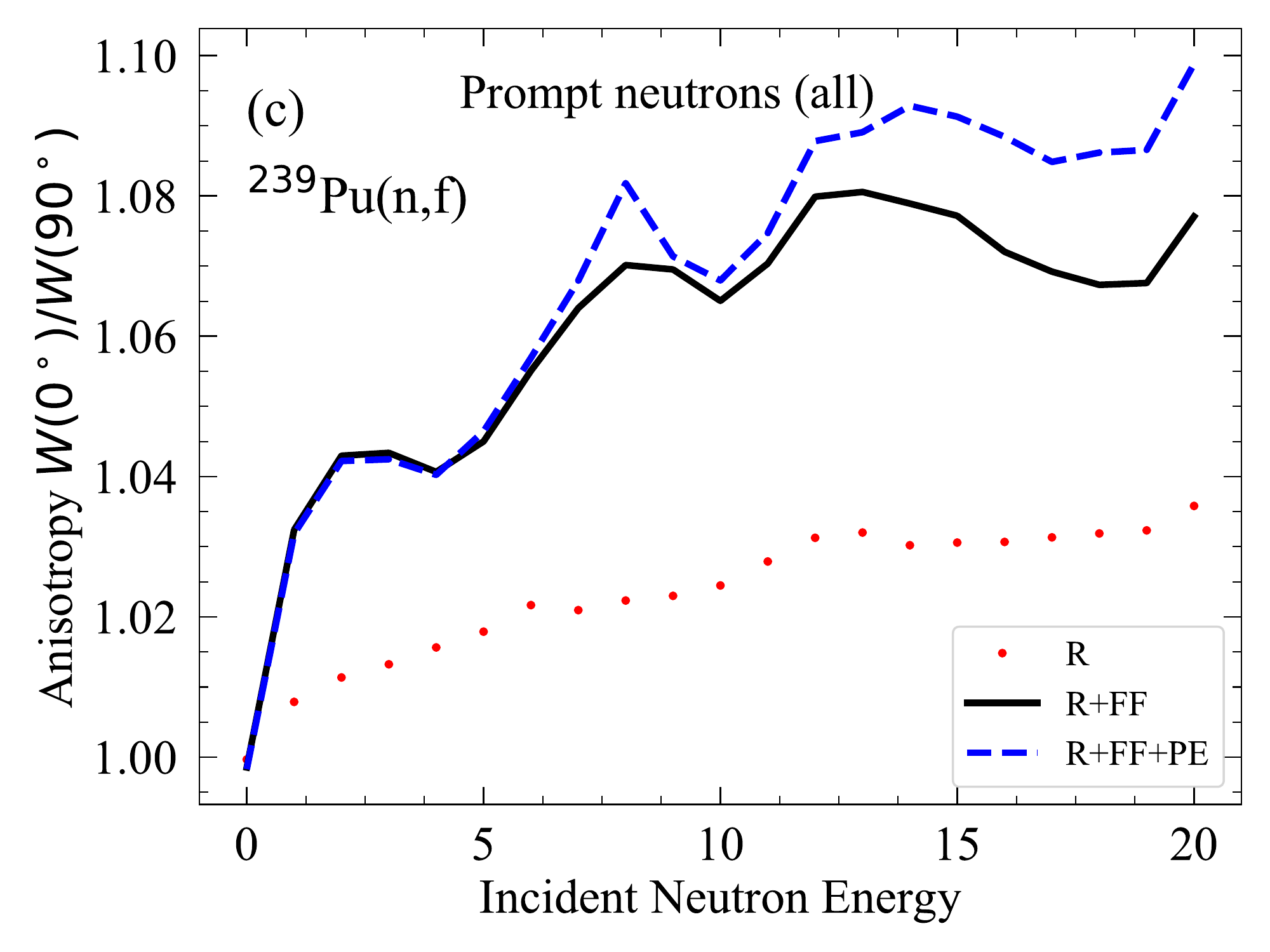} 
\caption{(color online) Anisotropy for (a) fission fragments, (b) prompt neutrons without pre-fission neutrons, and (c) all prompt neutrons as a function of incident neutron energy for different sources of angular anisotropy.  In each subplot, the red dotted line shows the effect due to the kinematic recoil (labeled R), the black solid line shows the joint effect of both the kinematic recoil and the fission fragment angular distribution (R+FF), and the blue dashed line depicts the anisotropy when the pre-equilibrium angular sampling is added to the first two effects (R+FF+PE), for $^{239}$Pu(n,f).  Note the difference in the y-scale across the three panels, especially for (a) compared to (b) and (c).}
\label{fig:FFAnisotropy}
\end{figure}

We now present results for the case of the neutron-induced fission on $^{239}$Pu for incident energies up to 20 MeV.  Because the sources of anisotropy can become entangled and more difficult to understand when we consider them together, we first examine, in sequence, the compounding effects due to the recoil of the target, the anisotropic distribution of the fission fragments, and then the angular sampling of the pre-equilibrium neutrons.  Figure \ref{fig:FFAnisotropy}(a) shows the compounding effect on the fission fragment angular distribution when each of these three effects is added.  The red dotted line shows the effect due to the recoil of the target from momentum conservation of the interaction with the beam, which can be up to 5\% at 20 MeV.  To this, the fission fragment anisotropy due to the transition states is added (black solid line).  Finally, the angular sampling of the pre-fission neutrons is included (blue dashed line).  As expected, there is very little change in the fragment angular distribution when the pre-equilibrium neutrons are sampled from an anisotropic distribution compared to the isotropic one.

The angular anisotropy of the emitted neutrons is lower compared to the fission fragment anisotropy.  The same additions as in Fig. \ref{fig:FFAnisotropy}(a) are shown in panel (b) but for the angular distributions of the neutrons emitted from the fission fragments, without pre-fission neutrons included.  Because the neutrons are emitted isotropically in the center-of-mass frame of each fission fragment, the values for the anisotropy are reduced.  Although the pre-equilibrium neutrons amount to less than 20\% of the pre-fission neutrons, there is a noticeable boost to the angular distribution at forward angles when these events are included.  The forward-peaked angular distribution translates to an increase in the anisotropy, as seen in the difference between the black solid and blue dashed lines in Fig. \ref{fig:FFAnisotropy}(c).  

\begin{figure}
\centering
\includegraphics[width=0.45\textwidth]{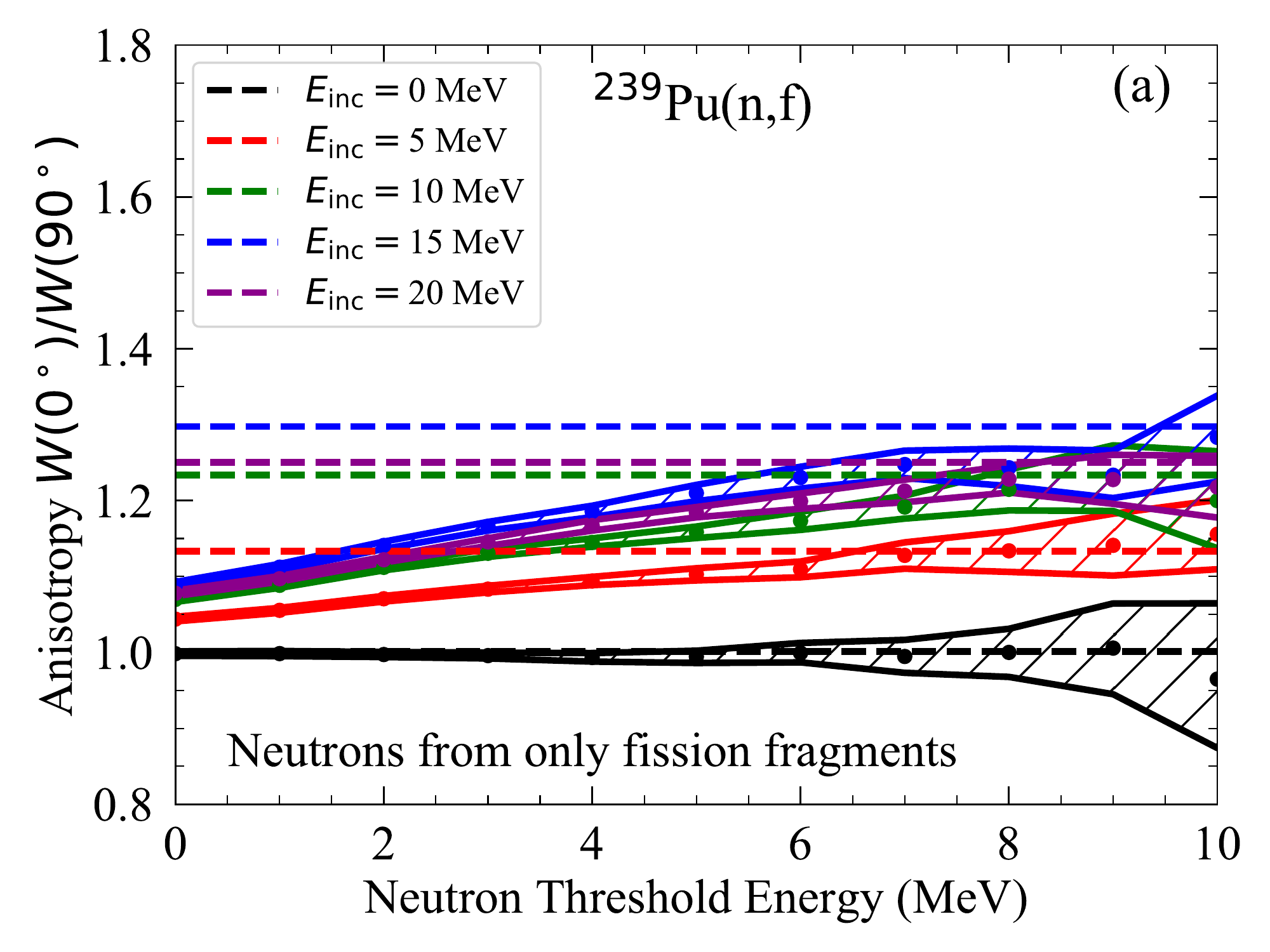} \\
\includegraphics[width=0.45\textwidth]{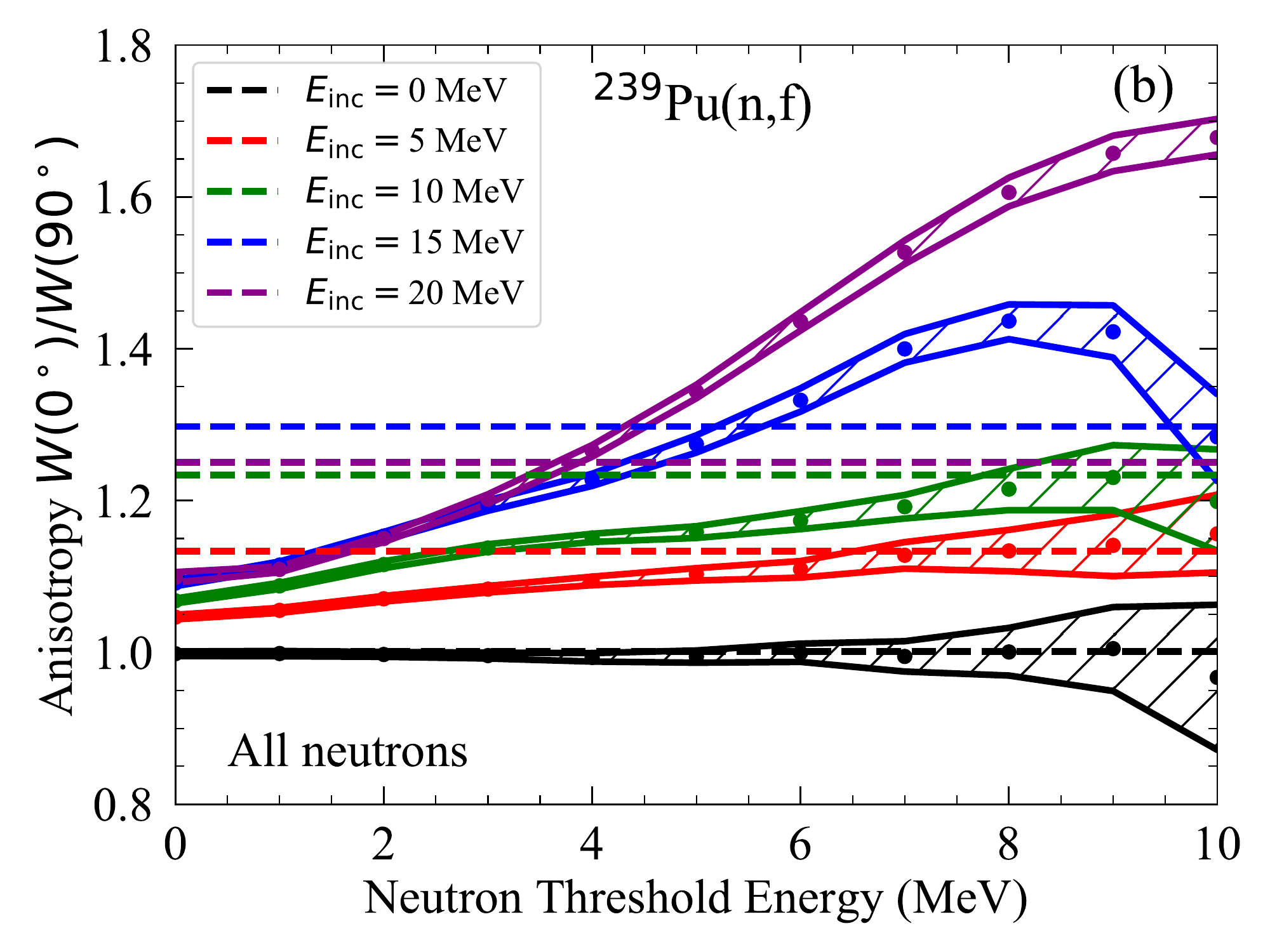} 
\caption{(color online) Neutron anisotropy as a function of threshold energy (filled circles) compared to the fission fragment anisotropy (dashed lines in the corresponding colors) (a) only neutrons emitted from the fission fragments are considered and (b) pre-fission neutrons are included in addition.}
\label{fig:nEth}
\end{figure}

We can define a lower bound on the outgoing neutron energy, the neutron threshold energy, $E_\mathrm{thres}$.  As $E_\mathrm{thres}$ is raised, the magnitudes of the neutron anisotropies are enhanced.  In Fig. \ref{fig:nEth}, we show the neutron anisotropy as a function of $E_\mathrm{thres}$ for several incident neutron energies, indicated by the different colored symbols.  Panel (a) shows the neutron anisotropy when pre-fission neutrons are removed from the analysis, and the neutron anisotropy when they are included is shown in panel (b).  In both panels, the hashed areas give the uncertainties coming from the fit to Legendre polynomials, as described in Sec. \ref{sec:extractA}.  When pre-fission neutrons are not considered, as in Fig. \ref{fig:nEth}(a), we see that the neutron anisotropy trends toward the fission fragment anisotropy as $E_\mathrm{thres}$ increases.  Although this picture is complicated by the addition of pre-fission neutrons, there appear to be strong correlations between the neutron anisotropy and $E_\mathrm{thres}$ that could be used to extract information about the fission fragment anisotropy.

%The results shown here are for $^{239}$Pu(n,f), but the same studies were also performed for $^{233,234,235,238}$U(n,f), $^{241}$Pu(n,f) and $^{237}$Np(n,f).  The trends seen for the case shown here were the same for each of the reactions studied, although we see some discrepancies when larger values of the fission fragment anisotropy are sampled.  This will be discussed further in the next section.

%%%%%%%%%%%%%%%%%%%%%%%%%%%%%%%%%%%%%%%%%%%%%%%%%%%%%%%%%%%%%%%%%%%%%%%%%%%%%%%%%%%%%%%%%%%%%%%%%%%%%%%%%%%%%%%%%

%\section{Discussion}
%\label{sec:discussion}

%\begin{itemize}
	%\item Anisotropy of the neutrons (caused by the fission fragments) is washed out compared to the anisotropy of the fission fragments - as seen in the angular distributions (also skewed because of the boost), do we have an analytical solution for this?
	%\item Neutron anisotropy is more prominent for higher energy neutrons (boosted more in the forward direction)
	%\item Can this be seen experimentally (especially with more limited angular coverage - $\approx$35/70 degrees instead of 0/90 degrees)?  Input from Chi-Nu folks
	%\item Discuss the pre-equilibrium neutrons as well
	%\item Does the anisotropy of the neutrons track the anisotropy of the fission fragments as a function of incident neutron energy, accounting for all of these details?  \IB{Only up to the emission of pre-equilibrium neutrons.}
	%\item How does this change with emitted neutron threshold energy?
	%\item Are the results different for the three fissioning systems?  \IB{Similar it seems, but more analysis is needed}
%\end{itemize}

\begin{figure}
\centering
\includegraphics[width=0.45\textwidth]{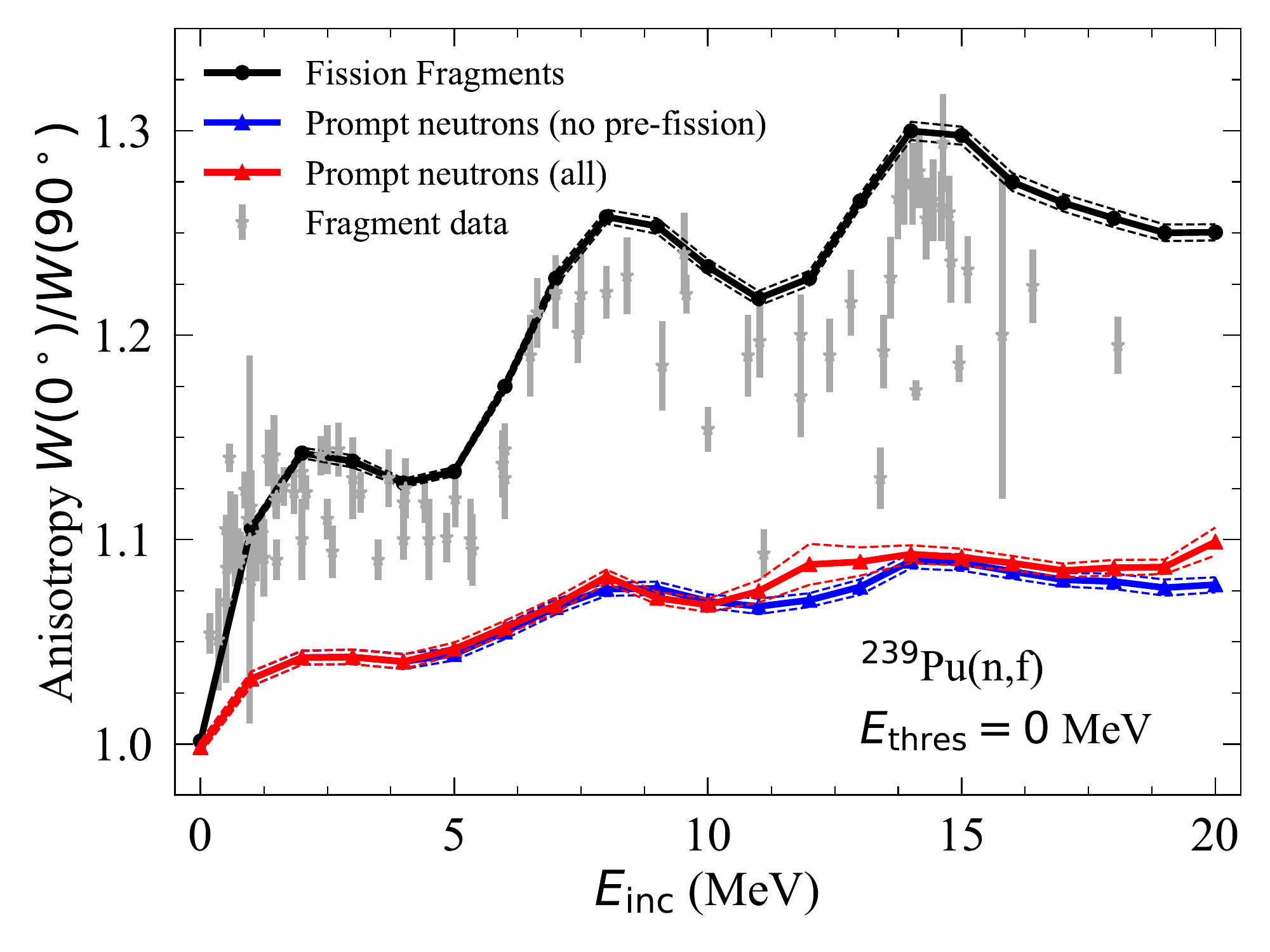}
\caption{(color online) Neutron anisotropy as a function of incident energy without (blue) and with (red) pre-fission neutrons included, compared to the experimental data (grey) that were used to construct the spline fit for {\CGMF} and resulting fission fragment anisotropy (black) when no neutron threshold energy is considered.  The dashed lines show the associated 95\% confidence bands of the Legendre polynomial fit.  The difference between the {\CGMF} calculation for the fission fragment anisotropy and the experimental data is discussed in the text.}
\label{fig:Eth0}
\end{figure}

As shown in Fig. \ref{fig:FFAnisotropy}, the emission of prompt neutrons dilutes the signature of the anisotropy coming from the fission fragments, as the neutrons are emitted isotropically in the center-of-mass frame of the fission fragments.  However, as seen in Fig. \ref{fig:nEth}(a), as the outgoing neutron energy increases, the neutron anisotropy approaches the fission fragment anisotropy.  Therefore, if neutrons with a high enough energy are measured, the fission fragment anisotropy could be extracted from the neutron anisotropy.  Of course, pre-fission neutrons complicate this picture, as in Fig. \ref{fig:nEth}(b), since they are emitted isotropically in the rest frame of the compound nucleus.  A hint of this difference when the pre-fission neutrons are included is seen in Fig. \ref{fig:Eth0} which shows the anisotropy of the fission fragments in black, prompt neutrons from the fragments in blue, and all neutrons in red, compared to the experimental data for the fission fragments, light grey circles.  Around $E_\mathrm{inc}=5$ MeV, where the second chance fission channel opens, very slight differences can be seen between the calculations of the neutron anisotropy with and without pre-fission neutrons, but these differences are primarily within the uncertainty bands from each calculation.  When the pre-equilibrium neutrons are first emitted, around $E_\mathrm{inc}=12$ MeV, the difference between the two neutron anisotropies becomes more visible, as some of the pre-fission neutrons are now emitted preferentially in the direction of the incident neutron beam.  

\begin{figure}
\centering
\includegraphics[width=0.45\textwidth]{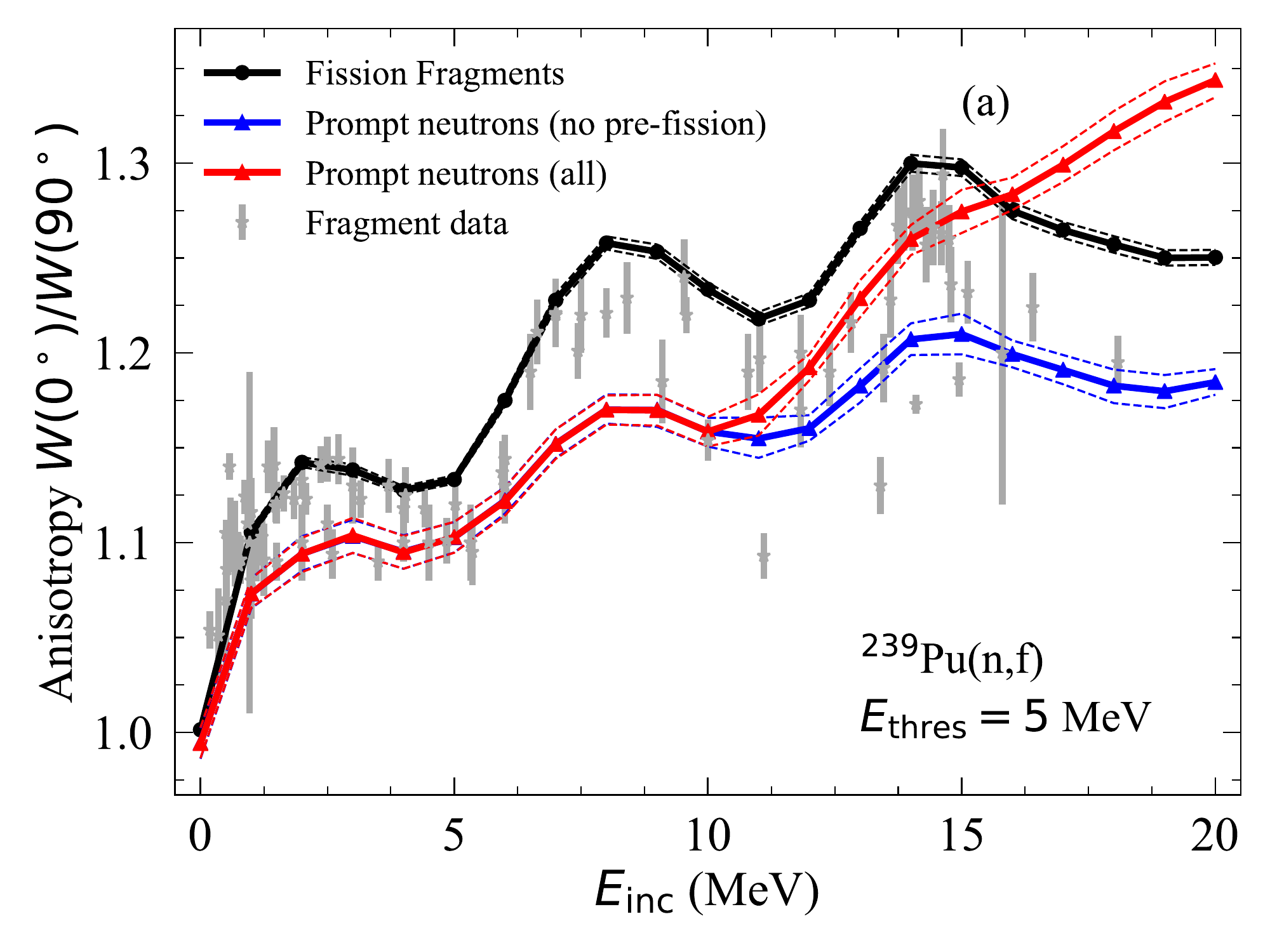} \\
\includegraphics[width=0.45\textwidth]{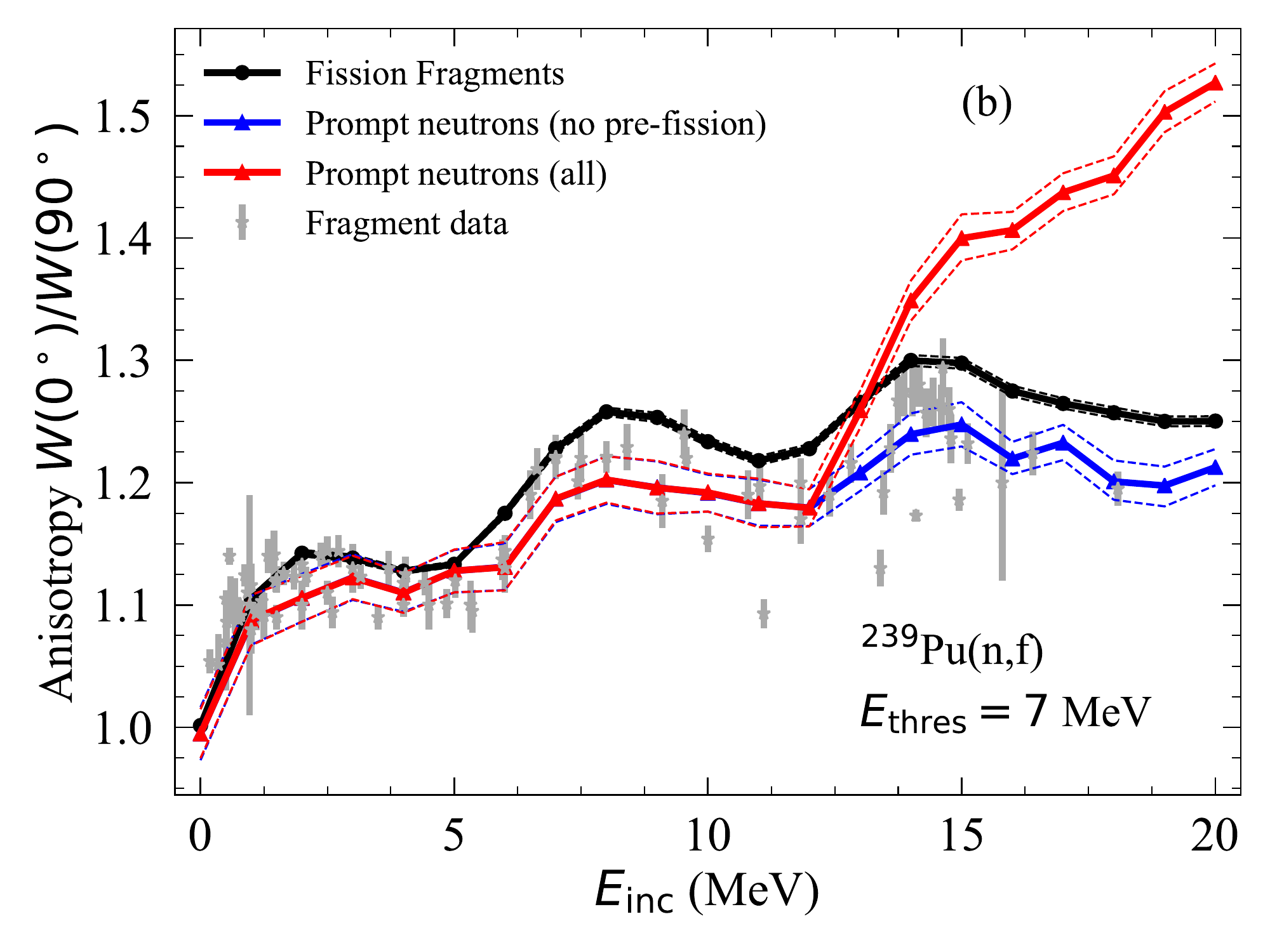} \\
\includegraphics[width=0.45\textwidth]{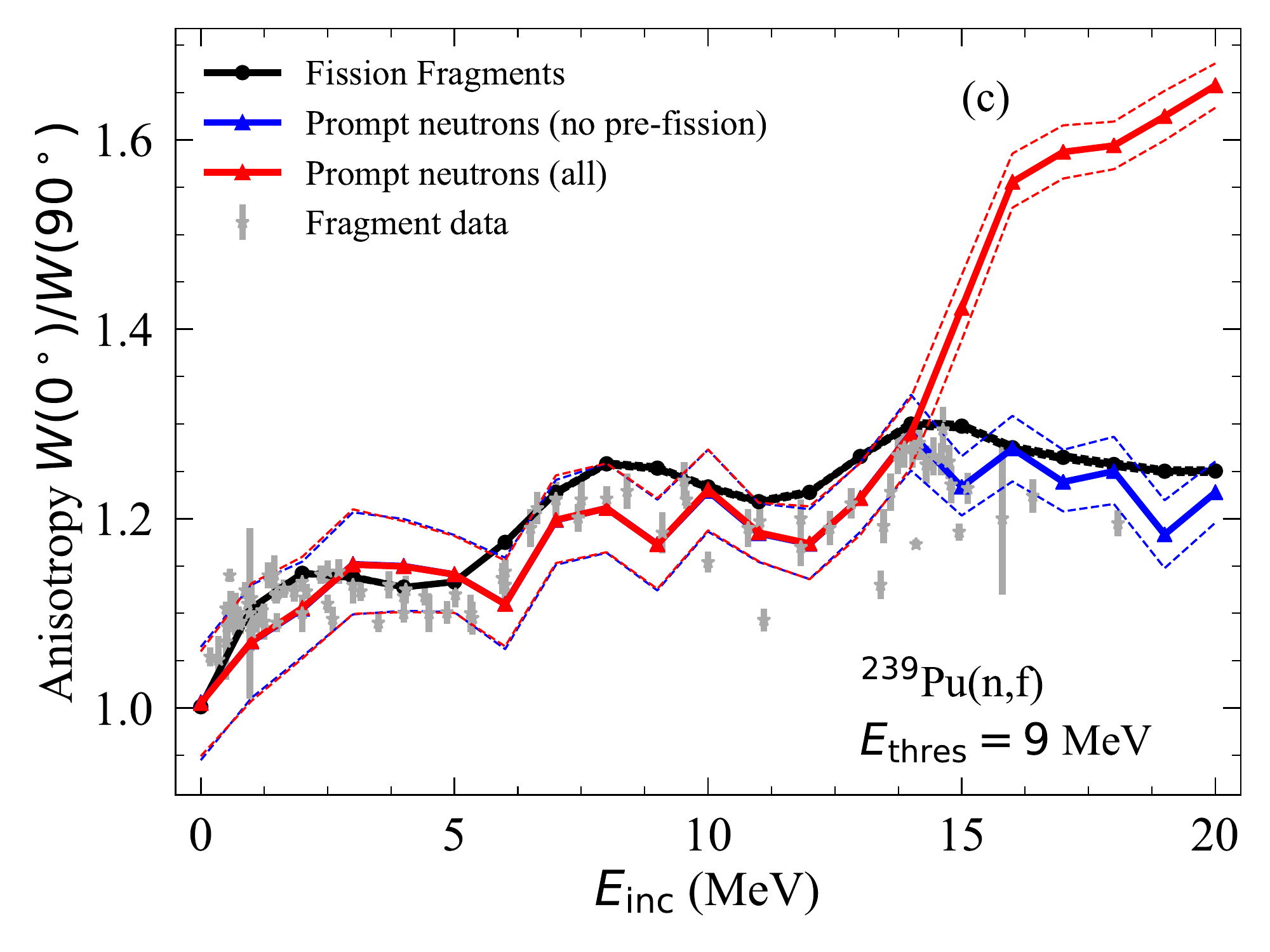} \\
\caption{(color online) Neutron anisotropy as a function of incident energy without (blue) and with (red) pre-fission neutrons included compared to the experimental data (grey) that was used to construct the spline fit for {\CGMF} and resulting fission fragment anisotropy (black).  The dashed lines show the associated 95\% confidence bands of the Legendre polynomial fit.  In (a) $E_\mathrm{thres}=5$ MeV, (b) $E_\mathrm{thres}=7$ MeV, and (c) $E_\mathrm{thres}=9$ MeV.}
\label{fig:239Pu}
\end{figure}

In most cases, the most energetic neutrons are those that are emitted along the direction of the fission fragments, as they gain the strongest kinematic boost from the motion of the fragments.  Therefore, as can be seen in the differences between panels (a), (b), and (c) in Fig. \ref{fig:239Pu}, as the energy threshold for the neutrons increases, the neutron anisotropy is more aligned with the anisotropy of the fission fragments.  When $E_\mathrm{thres}=7$ MeV, Fig. \ref{fig:239Pu}(b), the neutron anisotropy and associated uncertainties from the fit to the Legendre polynomials covers the spread in the experimental data, with 62\% of the data falling within the confidence bands for the neutron anisotropy when experimental uncertainty is taken into account.  We do see that the fission fragment anisotropy extracted from {\CGMF}, black solid lines in Fig. \ref{fig:239Pu}, is somewhat higher than the experimental anisotropy, due to the kinematic boost from the incident neutron beam.  The correction for the neutrons from the recoil of the target is less than 4\%.

Raising the neutron threshold energy pushes the effect of the pre-fission neutrons to higher incident energies.  For $E_\mathrm{thres}=5$ MeV, as in Fig. \ref{fig:239Pu}(a), pre-fission neutrons only begin to have an effect when the incident neutron energy is above 10 MeV.  As the threshold energy is increased, the pre-fission neutrons have an increasingly smaller effect on the overall neutron anisotropy; the difference in the anisotropy is only seen for higher incident neutron energies.  The pre-fission neutron energy spectrum is typically peaked around 2 MeV and drops off at higher outgoing neutron energies.  However, around $E_\mathrm{inc}\sim 9-10$ MeV, the pre-fission neutrons become more numerous than the neutrons emitted from the fission fragment and dominate the anisotropy when only higher neutron energies are considered.   As the neutron threshold energy continues to increase, the emission of high-energy pre-fission neutrons would not leave the residual compound with enough energy to fission.  Thus, the neutrons with the highest energies are emitted from the fission fragments, in the same direction as the fragments.  The trade-off here is that fewer neutrons are emitted with these highest energies, which is clear from the growing uncertainty bands between each of these energy thresholds.  

\begin{figure}
\centering
\includegraphics[width=0.45\textwidth]{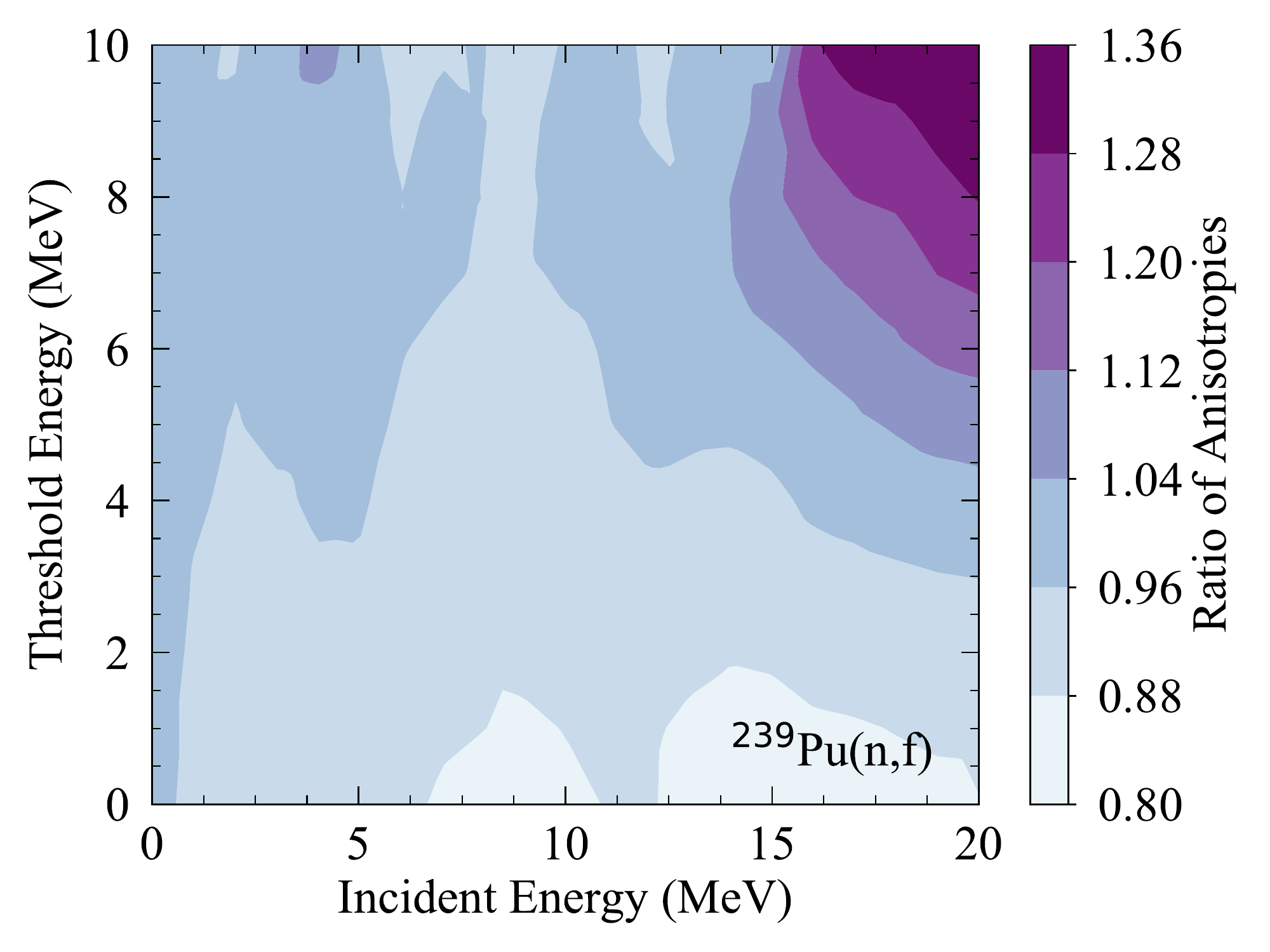}
\caption{(color online) Ratio between the neutron anisotropy (all prompt neutrons) and the fission fragment anisotropy as a function of the incident neutron energy and neutron threshold energy for $^{239}$Pu(n,f).}
\label{fig:ACorrection}
\end{figure}

We can then calculate a correlation factor between the neutron anisotropy and the fission fragment anisotropy, defined as the ratio of the neutron anisotropy (including pre-fission neutrons) to the fission fragment anisotropy at every incident and threshold neutron energy.  In Fig. \ref{fig:ACorrection}, this ratio is plotted as a function of incident neutron energy and threshold neutron energy.  Thus, knowing the incident and outgoing neutron energies and measuring the neutron anisotropy, the fission fragment anisotropy could be extracted with this correlation factor.  It is worth noting that the correlation factor rises above one at the highest incident energies and neutron threshold energies due to the inclusion of the pre-fission neutrons, consistent with what is shown in Fig. \ref{fig:239Pu} in red.

We also performed these studies for neutron-induced fission on $^{233,234,235,238}$U, $^{237}$Np, and $^{241}$Pu, all available in {\CGMF}.  In all cases, we found very similar results to what was shown here for $^{239}$Pu(n,f).  The most notable difference was for even isotopes of uranium, which were the only even-even targets studied here.  Differences in the spin of the target and the available transition states can lead to large differences in the fission fragment anisotropy between even-even and even-odd target nuclei \cite{Simmons1960,Mueller2012}.  An increase of anisotropy is seen for all targets at the opening of the higher multi-chance fission channels, due to the population of few transition states in residual fissioning nucleus with less excitation energy.  The density of states in the residual nucleus depends on the odd/even character of the nucleus.  For these isotopes, the values of the anisotropy of the fission fragments could reach nearly to 2, significantly larger than any of the values shown here for $^{239}$Pu(n,f).  When the fragment anisotropy was this large, the neutron anisotropy did not reach those values, regardless of $E_\mathrm{thres}$ (within the limits of the statistics of our calculations).  When the fragment anisotropy is high, increasing the energy threshold for the neutrons has only a small effect on the resulting neutron anisotropy.  For this case, the neutrons seem to already be boosted in the direction of the fission fragments.  Increasing the neutron threshold energy would then have a smaller effect on the resulting neutron anisotropy, as fewer neutrons are being excluded based on these cuts.  Further studies on this feature should be performed.

%\subsection{Comparison to experimental data}
%\label{sec:data}

Recently, several measurements of the angles and energies of prompt neutrons from the fission of major actinides have been made with the Chi-Nu detector array at LANL.  The Chi-Nu experimental setup consists of two detector arrays used in separate measurements:  a 22-detector Li-glass array to measure outgoing neutrons with energy from 0.01 -- 1.59 MeV and a 54-detector liquid scintillator array to measure outgoing neutrons from approximately 0.89 -- 10 MeV.  The spectra from these detectors are combined to form prompt fission neutron spectrum (PFNS) measurements from 10 keV to 10 MeV for incident neutron energies from 1 -- 20 MeV.  More details can be found in \cite{Devlin2018,Kelly2018a,Kelly2018b}.  Prompt neutrons from neutron-induced fission on $^{238}$U were measured with the Chi-Nu liquid scintillator array to study the PFNS up to 20 MeV incident neutron energy.  The Chi-Nu liquid scintillator array covers nine outgoing neutron angles from 30 -- 150$^\circ$ in the lab frame, allowing for a comparison between the {\CGMF} calculations of the neutron anisotropies to those measured by Chi-Nu.

In order to exclude many sources of systematic error from the experiment, the neutron anisotropies from Chi-Nu are presented as a ratio of experimental data at 30$^\circ$ and 90$^\circ$, normalized to the lowest incident neutron energy range,
\begin{equation}
\rho(E_\alpha) = \frac{\sum \limits _i C(E_i,E_\alpha,30^\circ)/C(E_i,E_\mu,30^\circ)}{\sum \limits _i C(E_i,E_\alpha,90^\circ)/C(E_i,E_\mu,90^\circ)}.
\label{eqn:ChiNuRatio}
\end{equation}

\noindent$E_i$ are the outgoing neutron energies, $E_\alpha$ and $E_\mu$ are incident neutron energies, where $E_\mu$ is the same for each of the ratio calculations and is taken to be the lowest measured incident neutron energy, and $C(E_i,E_\alpha,\theta)$ is the number of foreground counts in the detector at angle $\theta$ for $E_i$ and $E_\alpha$.  This formulation allows many sources of uncertainty to cancel or cancel approximately, including the number of fissions, the neutron detector efficiencies, and the neutron scattering corrections, leaving primarily statistical uncertainties.  The ratio resulting from Eq. (\ref{eqn:ChiNuRatio}) is intrinsically a shape quantity as a function of incident neutron energy that contains a reflection of the fission fragment anisotropy.  Because this shape is, by default, scaled to have $\rho(E_\alpha)=1$ at the lowest incident energy ($E_\mathrm{inc}=1.54$ MeV), and there is no measurement at thermal energies where the ratio should be one, this shape was scaled to reproduce the magnitude of the {\CGMF} calculations from $6 - 14$ MeV incident neutron energy.  

\begin{figure}
\centering
\includegraphics[width=0.45\textwidth]{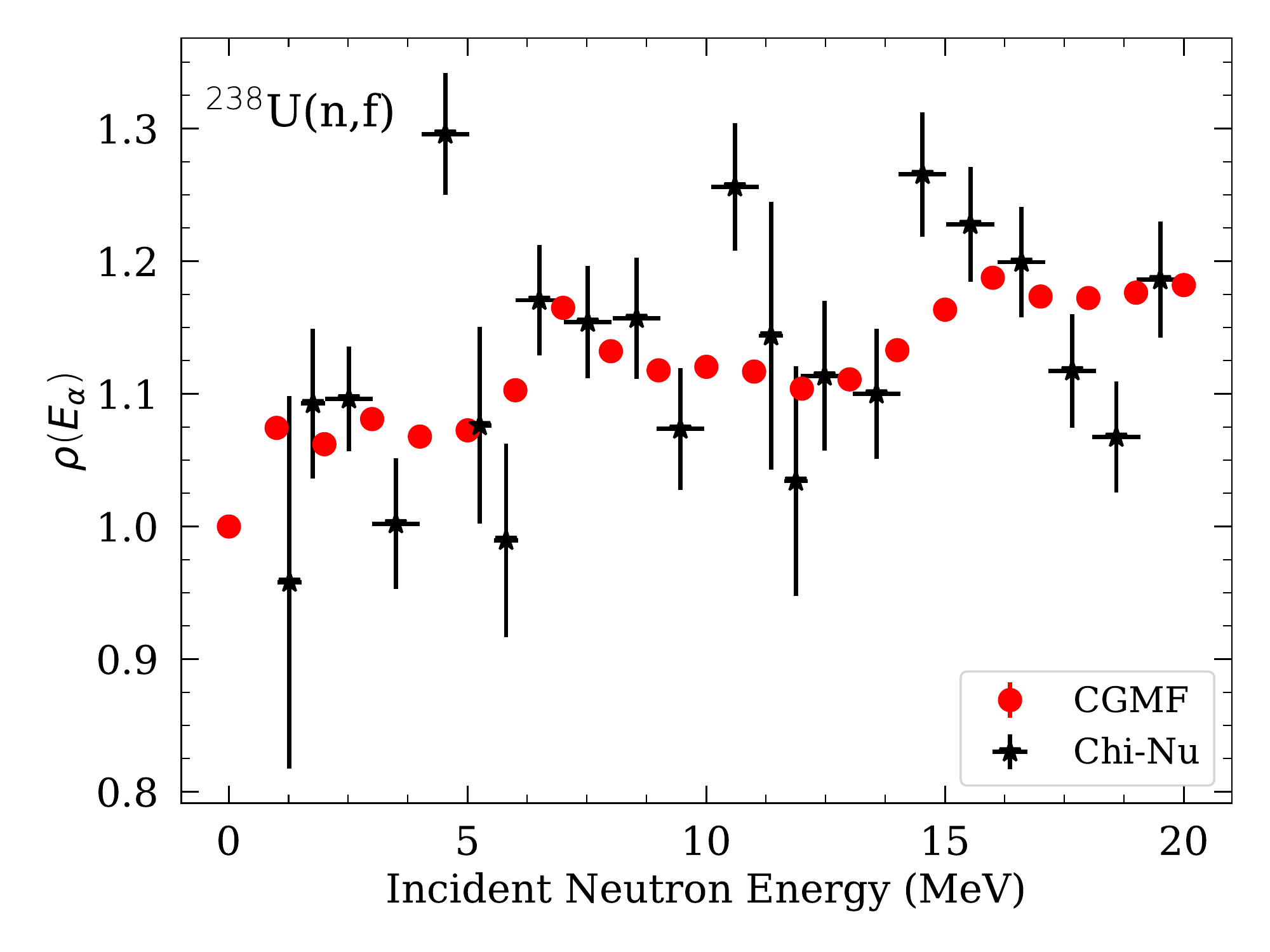}
\caption{Comparison between {\CGMF} calculations, red circles, and preliminary Chi-Nu data, black stars, for $\rho(E_\alpha)$ as given in Eq. (\ref{eqn:ChiNuRatio}) for $^{238}$U(n,f).}
\label{fig:U238ChiNuComp}
\end{figure}

The same ratio can be calculated from the {\CGMF} histories and compared to the measurement from Chi-Nu.  This comparison is shown in Fig. \ref{fig:U238ChiNuComp}.  For the best trade-off between statistics and high incident neutron energies where the neutron anisotropy most closely tracks the fission fragment anisotropy, the sum in Eq. (\ref{eqn:ChiNuRatio}) is taken from $E_i = 5 - 10$ MeV.  The data from Chi-Nu are shown as the black filled stars, compared to the {\CGMF} calculation of Eq. (\ref{eqn:ChiNuRatio}) as the red filled circles.  The {\CGMF} calculations are scaled such that $\rho(E_\alpha)=1$ for thermal neutrons, and Chi-Nu was scaled to match the relative magnitude of the {\CGMF} calculations to be able to compare the shapes of the two ratios.

The shapes of the distributions from Chi-Nu and {\CGMF} are mostly consistent within the experimental uncertainties.  The experimental uncertainties are large compared to those of the calculation, which are only statistical and do not account for model defects.  In the calculated and measured ratios, there is a rise up to $E_\mathrm{inc} \sim 2$ MeV, followed by a flattening before the second-chance fission channel opens around incident energies of 6 MeV.  Both ratios flatten off again, and then there is another increase in $\rho(E_\alpha)$ where the third-chance fission channel opens, although there could be a disagreement between the incident neutron energy where this channel opens experimentally and in {\CGMF}.  The Chi-Nu experiment was not optimized to extract the neutron anisotropy, and a different set-up could be put in place to more accurately measure this quantity.  The {\CGMF} calculations could then be used to convert the experimental values of $\rho(E_\alpha)$ to fission fragment anisotropies and compare with data measured directly.%  This will be explored in future work.

%%%%%%%%%%%%%%%%%%%%%%%%%%%%%%%%%%%%%%%%%%%%%%%%%%%%%%%%%%%%%%%%%%%%%%%%%%%%%%%%%%%%%%%%%%%%%%%%%%%%%%%%%%%%%%%%%

\section{Conclusions}
\label{sec:conclusions}

%\begin{itemize}
	%\item Can we determine fission fragment anisotropy from prompt neutron anisotropy?  Yes
	%\item If so, to what extent (do pre-fission neutrons complicate the picture)?  Yes
	%\item Is this limited in incident energy or threshold energy?  Yes
%\end{itemize}

In summary, angular distribution sampling for both the fission fragment anisotropy and the pre-equilibrium neutrons were recently introduced into the Monte Carlo fission code {\CGMF}.  Experimental values for the fission fragment anisotropy were fitted using cubic splines while the pre-equilibrium neutron angular distributions were calculated with the semi-direct model of Feshbach, Kerman, and Koonin and fitted as a function of neutron incident energy and compound excitation energy using Kalbach systematics.  In itself, the inclusion of these effects is important to interpret experimental results from detector setups that measure neutron angles and energies, such as the Chi-Nu array.  We also found that high-energy neutrons track the fission fragment anisotropy relatively well, and this congruence improves as the energy threshold of the neutrons is increased.  This was studied for neutron-induced fission reactions on $^{233,234,235,238}$U, $^{237}$Np, and $^{239,241}$Pu.

Excluding pre-fission neutrons, for a threshold energy of 7 MeV for the outgoing neutrons, the uncertainty of the theory coming from the Legendre polynomial fit to the {\CGMF} calculations is similar to the spread in the available data.  When the pre-fission neutrons are included, the neutron anisotropy does not change until after the opening of third-chance fission.  This difference in the neutron anisotropy when not including and including pre-fission neutrons is only seen at increasingly higher incident neutron energies if the neutron threshold energy is increased, but statistics become significantly worse.  A correlation factor, such as the one calculated here with {\CGMF}, could be used to extract the fission fragment anisotropy from the neutron anisotropy or vise-versa depending on the incident and outgoing threshold neutron energies, even when pre-fission neutrons are included in the analysis.  Here, we performed a preliminary comparison with $^{238}$U(n,f) data measured at Chi-Nu, using a ratio method which showed rather good agreement between the shapes of the {\CGMF} calculations and experimental measurement.

There are still other effects that we have not taken into account in this study that could also contribute to neutron anisotropy, including the angular distribution of the neutrons in the center-of-mass frame of the fission fragments, the emission of neutrons from fission fragments that are not fully accelerated, and the emission of scission neutrons.  These effects should be studied theoretically and experimentally and included in the model.
 
%%%%%%%%%%%%%%%%%%%%%%%%%%%%%%%%%%%%%%%%%%%%%%%%%%%%%%%%%%%%%%%%%%%%%%%%%%%%%%%%%%%%%%%%%%%%%%%%%

%\begin{itemize}
%	\item Funding - CNLS, NA22
%\end{itemize}
\begin{acknowledgements}
The authors would like to thank Matt Devlin and John O'Donnell for their feedback on the comparison to the Chi-Nu data and Toshihiko Kawano for his pre-equilibrium neutron calculations and helpful discussions about those calculations.  This work was performed under the auspice of the U.S. Department of Energy by Los Alamos National Laboratory under Contract 89233218CNA000001 and was supported by the Office of Defense Nuclear Nonproliferation Research \& Development (DNN R\&D), National Nuclear Security Administration, U.S. Department of Energy.  We gratefully acknowledge the support of the U.S. Department of Energy through the LANL/LDRD Program and the Center for Non Linear Studies.
\end{acknowledgements}

%%%%%%%%%%%%%%%%%%%%%%%%%%%%%%%%%%%%%%%%%%%%%%%%%%%%%%%%%%%%%%%%%%%%%%%%%%%%%%%%%%%%%%%%%%%%%%%%%

%\appendix

%%%%%%%%%%%%%%%%%%%%%%%%%%%%%%%%%%%%%%%%%%%%%%%%%%%%%%%%%%%%%%%%%%%%%%%%%%%%%%%%%%%%%%%%%%%%%%%%%

\bibliography{aniso}

\end{document}